\documentclass{PoS}

\def\Sw{{\em Swift}}
\def\Ss{{\em Swift}~}
\def\Ep{$E_{\rm pk}$}
\def\Eps{$E_{\rm pk}$~}

\def\gsim{\mathrel{\raise.5ex\hbox{$>$}\mkern-14mu
             \lower0.6ex\hbox{$\sim$}}}
\def\lsim{\mathrel{\raise.3ex\hbox{$<$}\mkern-14mu
             \lower0.6ex\hbox{$\sim$}}}

\title{The Statistics of the Prompt-to-Afterglow GRB Flux Ratios and the
Supercritical Pile GRB Model
 }

\ShortTitle{Statistics of GRB Prompt/Plateau Flux Ratios}

\author{\speaker{D. Kazanas}\thanks{Work supported by {\em Fermi} GI funds}\\
        NASA/GSFC, Code 663, Greenbelt, MD 20771\\
        E-mail: \email{Demos.Kazanas@nasa.gov}}

\author{J. L. Racusin\\
        NASA/GSFC, Code 661, Greenbelt, MD 20771\\
        E-mail: \email{judith.racusin@nasa.gov}}

\author{J. Sultana\\
        Mathematics Department, Faculty of Science,
University of Malta, Msida MSD2080 Malta\\
        E-mail: \email{joseph.sultana@um.edu.mt}}

\author{A. Mastichiadis\\
        Department of Physics, University of Athens,
Panepistimiopolis, GR 15783, Zografos, Greece\\
        E-mail: \email{amastich@physics.uoa.gr}}

\abstract{We present the statistics of the ratio, ${\mathrm R}$, between the prompt and
afterglow "plateau" fluxes of GRB. This we define as the ratio
between the mean prompt energy flux in the {\em Swift} BAT and the {\em
Swift} XRT, immediately following the steep transition between
these two states and the beginning of the afterglow stage referred
to as the "plateau". Like the distribution of other GRB observables, the
histogram of ${\mathrm R}$ is close to log-normal, with maximum at
${\mathrm R = R}_{\rm m} \simeq 2,000$, FWHM of about 2 decades and with the
entire distribution spanning about 6 decades in the value of ${\mathrm R}$. We
note that the peak of the distribution is close to the
proton-to-electron mass ratio $({\mathrm R}_{\rm m} \simeq m_p/m_e = 1836)$, as
proposed by us earlier, on the basis of a specific model for the conversion of
the GRB blast wave kinetic energy into radiation, before any similar
analysis were made. It therefore appears that, in addition to the values of
the energy of peak luminosity ${E_{\rm pk}\sim m_{e} c^2}$, GRB present us
with one more quantity with an apparently characteristic value. The fact that
the values of both these quantities (i.e. $E_{\rm pk}$ and ${\mathrm R}$) comply with
those implied by the same specific model devised to account for an altogether
different issue, namely the efficient conversion of the GRB blast wave kinetic
energy into radiation, argues favorably for its underlying assumptions.\ }

\FullConference{10th INTEGRAL Workshop: "A Synergistic View of the High Energy Sky" - Integral2014,\\
		15-19 September 2014\\
		Annapolis, MD, USA }

\begin{document}

\section{Introduction}

The nature of Gamma-Ray Bursts (GRB), short duration ($\sim 10^{-2}
- 10^2$ s), $\gamma-$ray emission events at cosmological distances,
remains enigmatic despite much observational and theoretical
progress over the past twenty five years. Furthermore, they appear
to present us with novel puzzles each time improvements in
instrumentation allow for significant improvement in measuring their
properties. One such example is the shape of their afterglow light
curves, which the \Ss observations showed to be vastly different
from those predicted
by the early theoretical considerations, 
the subject of the present note.

The cosmological origin of GRB was essentially confirmed with the
discovery of their afterglows. This set the redshifts of their host
galaxies, determined their distances to be cosmological and their
luminosities in the range of $\sim 10^{50}-10^{54}$ erg/s. Their
implied high specific intensities, then, argued for emission by the
plasma of relativistic blast waves  (RBW) of Lorentz factor (LF)
$\Gamma \gsim 200$ \cite{rees92}, a fundamental tenet of the physics
of these events. However, while the association of GRB with RBW is
undisputable, there are still serious gaps in our fundamental
understanding of their radiation emission: Ignoring for the moment
the origin of flows with $\Gamma \gsim 200$, a RBW is an inertial
agent, distributing the postshock energy in proportion to the swept
particles' mass. So, the electrons, the only particles that can
radiate efficiently, carry only a fraction ($\eta \sim 1/2000 \sim
m_e/m_p$) of the shock's energy. Because such a small efficiency
would lead to unrealistically large GRB energies and luminosities,
the issue is resolved by {\em assuming} that the protons and
electrons share equally the postshock energy. Another open issue
associated with the GRB prompt stage is the rather limited range of
the energy of peak luminosity, $E_{\rm pk} \sim 0.3$ MeV (extending
on occasions to a few MeV), intriguingly close to the electron rest
mass on the Earth frame, but not on the GRB rest frame, considering
the large Lorentz factors of their blast waves. This characteristic
energy is {\em the} defining attribute of the GRB prompt emission
since the energy of peak luminosity declines very fast as GRB
develop in time and enter their, less variable, afterglow stage.

Afterglows have been considered to be, ever since their discovery, a
distinct and separate phase of the GRB phenomenon. However, there
does not appear to be a formal criterion that determines the
transition from the prompt to the afterglow GRB phase.
Operationally, the prompt phase ends when the GRB flux drops below
the \Sw-BAT sensitivity threshold. On theoretical grounds, it is
generally considered that the prompt GRB phase terminates when the
RBW reaches its deceleration radius $R_D$, since, beyond this point
its LF begins to decrease ($R_D$ is the radius at which the RBW has
swept-up mass-energy $Mc^2 \simeq m_p c^2 nR_D^3 \simeq E/\Gamma^2$;
$E$ is total energy of the RBW, $n$ the circumburst density -
assumed to be constant - and $\Gamma$ its Lorentz factor). The
decrease of the GRB Lorentz factor for $R > R_D$ ($\Gamma \propto
R^{-3/2}$ for adiabatic evolution in a uniform density medium) leads
to a decrease in the GRB flux and to the energy of its peak
luminosity, $E_{\rm pk}$, which moves out of the \Sw-BAT range. The
afterglow emission, just as that of the prompt phase, is considered
in most models to be synchrotron radiation by shock accelerated
electrons, in rough equipartition with the protons of the postshock
region. Under these assumptions, one can compute for the GRB
afterglow stage the resulting spectrum and its time evolution. The
computation of the GRB X-ray flux evolution in time was first
performed for spherically symmetric outflows in \cite{sar98} and in
\cite{sar99} for jet-like outflows, assuming the spectrum of
electrons injected at the shock to be $dN/dE \propto E^{-p}, ~ p
\gsim 2$. The model X-ray light curves were, then, shown to be power
laws in time, $F_X \propto t^{-\alpha}, ~ \alpha \gsim 1$, with a
smooth transition between the prompt and afterglow GRB stages, in
broad agreement with the sparsely sampled afterglow light curves of
the pre--{\em Swift} era.

The launch of \Ss and its ability to follow closely the evolution of
GRBs from their prompt ($\gamma-$ray emitting) to the afterglow
(X--ray emitting) stages provided yet another set of unexpected,
puzzling facts, grossly inconsistent with expectations based on the
models described above \cite{tagliaferri05,nousek,zhang06,evans09}:
Instead of the predicted decrease of their X-ray flux as $L_X
\propto t^{-\alpha}, \alpha \gsim 1$, the decrease is much steeper
($\alpha \sim 3 - 6$), followed either ({\em i}$\,$) by a much
shallower section (referred to as the "plateau") ($\alpha \sim 0$),
which is succeeded for $t > T_{\rm brk} \sim 10^3-10^5$ sec by a
decline of $\alpha \simeq 1$ or ({\em ii}$\,$) by a more
conventional decline ($\alpha \simeq 1$) (these light curves exhibit
also occasionally large amplitude flares which we will not discuss
at present).

The unexpected afterglow light curves prompted a number of accounts
of their behavior within the standard GRB model [see \S 3.1.1 of
\cite{Z07} and references therein]. The most conventional such
account is that the steep decay represents high latitude (with
respect to the observer's line of sight) emission; while steeper
($\alpha \simeq 2$) than the law derived in \cite{sar98}, this is
still less steep than most commonly observed and certainly at great
odds with declines as steep as $t^{-6}$. Alternative explanations
attribute this to the form of the underlying electron distribution
\cite{kaz07,GS09} while \cite{PMP11}, by adjusting the maximum
energy $\gamma_{max}$ of the electron distribution, interpret the
steep decline segment as synchrotron emission by the fast cooling,
high energy cutoff of the electron distribution function and the
``plateau" segment to inverse Compton by its more slowly varying low
energy section.

The plateau segment, because it follows that of the very steep flux
decline, gives the impression of a distinct and completely separate
emission from that of the GRB prompt phase. For this reason, it was
proposed (not unreasonably) that it indicates the emergence of an
additional injection of energy by the GRB central source, separate
from that producing the shorter but brighter prompt emission
\cite{WillOBr}. Along these lines, the authors of \cite{Gompertz14}
even specify this additional injection to be due to the propeller
effect of an underlying magnetar which presumably powers the entire
burst. Anyway, these attempts to account for the behavior of the GRB
afterglow light curves were devised to model these specific
features, without any reference to, or consideration of, the broader
properties of the entire burst. Alternatively, several treatments
have focused instead on just the properties of the plateau phase
itself, leaving its origin unspecified. Of these we mention those of
\cite{Lei11} and \cite{Matzner12} who, by fitting the
spectro-temporal evolution of the afterglow plateau segment of
several GRB, conclude that this emission takes place {\em before}
the RBW has reached its deceleration radius $R_D$.

\begin{figure}[t]
\begin{center}$
\begin{array}{cc}
\includegraphics[trim=0in 0in 0in 0in,keepaspectratio=false,
width=3.0in,angle=-0,clip=false]{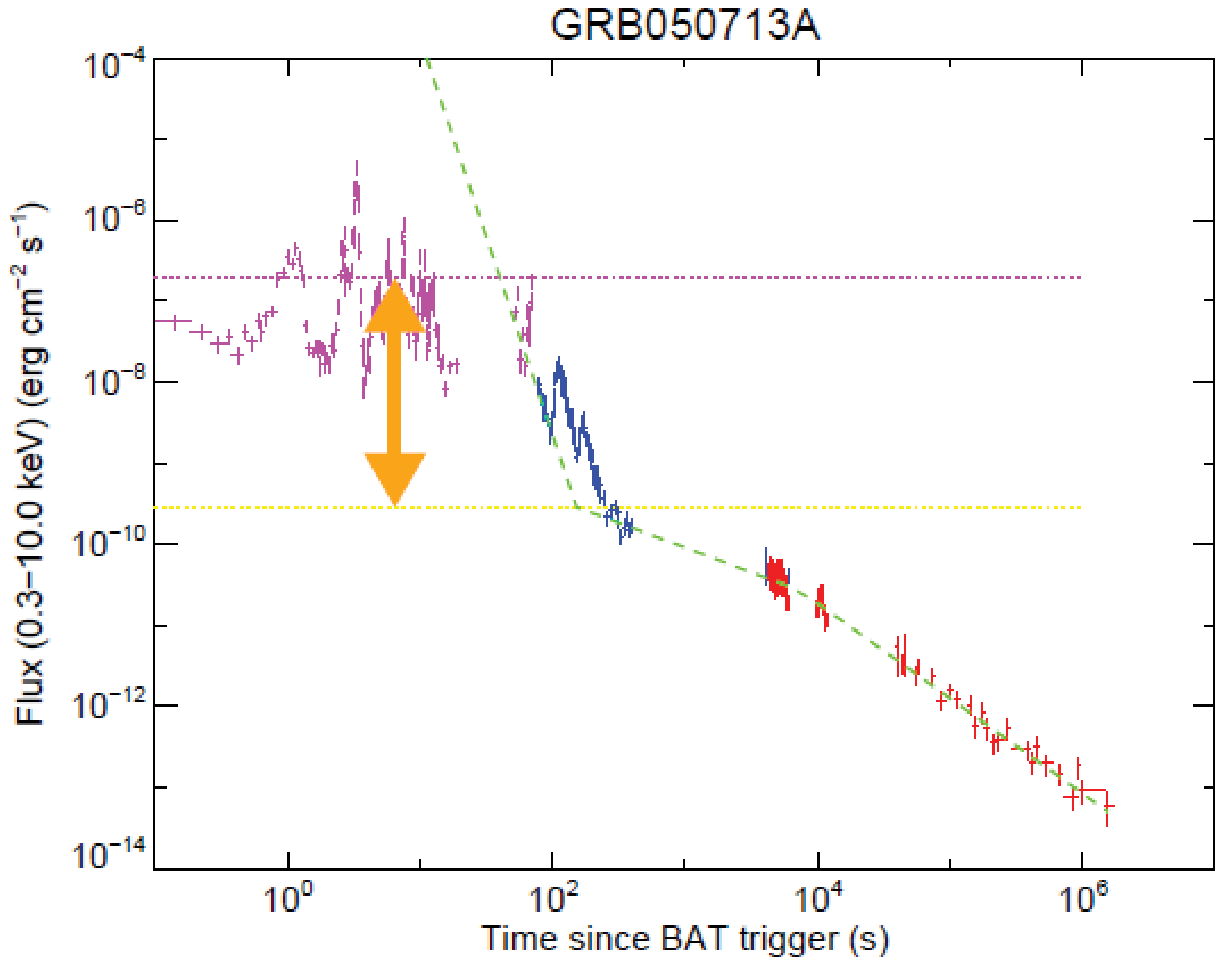}
&
\includegraphics[trim=0in 0in 0in 0in,keepaspectratio=false,
width=3.0in,angle=-0,clip=false]{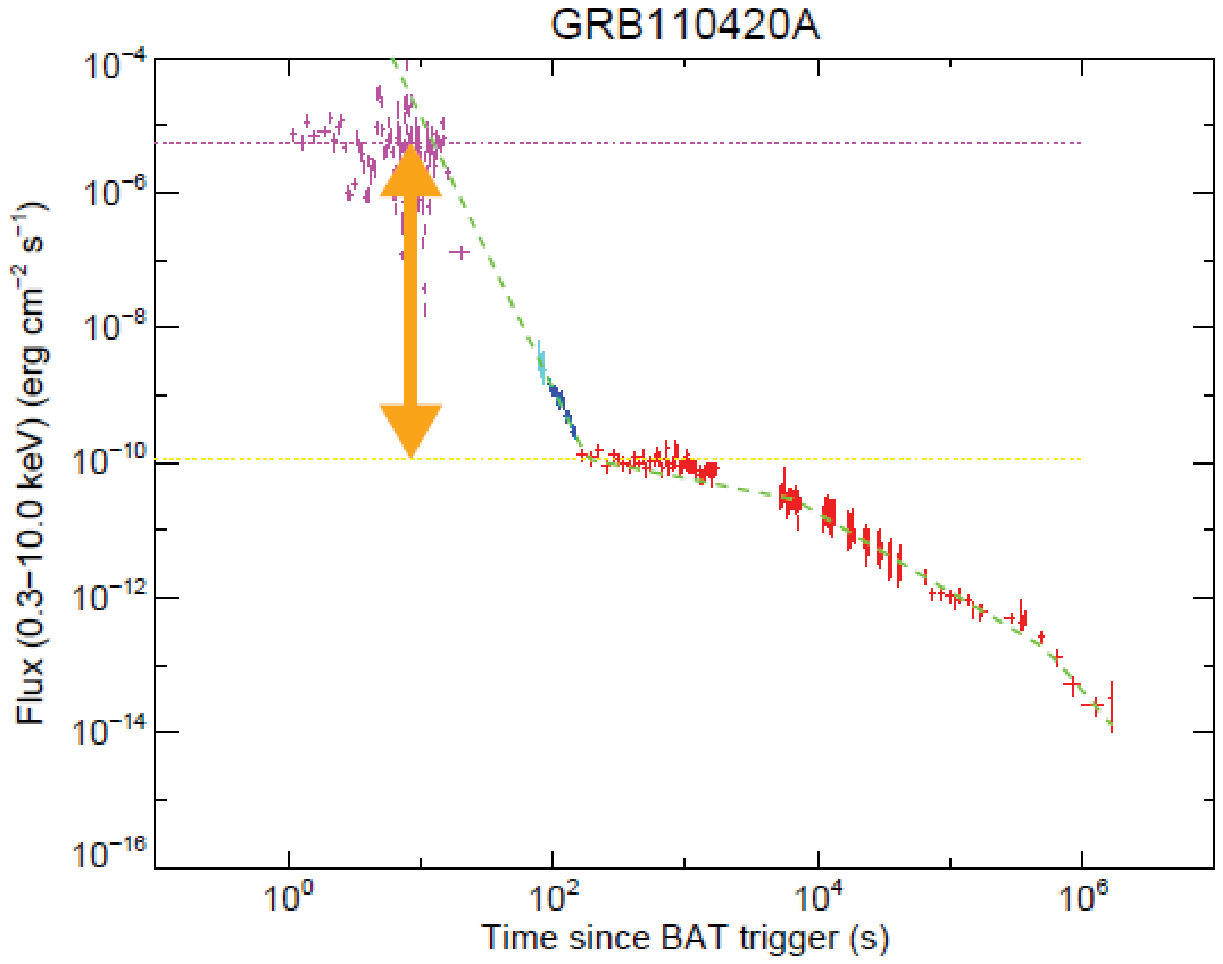}
\end{array}$
\end{center}
\vskip -20pt
\caption{\footnotesize  (a) The prompt to afterglow light curves of the gamma ray bursts
indicated on the figure. The prompt, transition and afterglow plateau stages are
apparent. The two dotted lines represent the mean prompt flux (top, purple line) and
afterglow (bottom, red line) fluxes involved in computing the flux ratios of these
two states. }
\label{fig:f1}
\end{figure}

\section{Afterglows in the Supercritical Pile Model (SPM)}

An altogether different approach to the afterglow evolution has been
that of \cite{SKM13}. This is different in that the afterglow
evolution, including all its details, is produced as an integral
part of the evolution of the entire burst, beginning with the
accelerating phase of the RBW  and continuing with its dissipation
and prompt emission, including also the correct value of the energy of
the GRB prompt phase emission, \Ep. The central notion of this model
is a radiative instability that converts the relativistic proton
energy behind the RBW forward shock to $e^+e^-$--pairs through the
$p \, \gamma \rightarrow p \, e^+e^-$ reaction; the pairs then
produce more synchrotron photons, which produce more pairs and so on
\cite{kaz02,mas06,mas08,mas09}. This instability requires that the
column of relativistic protons in the postshock region be larger
than a critical value, in direct analogy with a supercritical
nuclear pile (hence the nomenclature Supercritical Pile Model or
SPM); put simply, one cannot accumulate arbitrarily large columns of
relativistic protons for the same reason that one cannot accumulate
arbitrarily large amounts of U$^{235}$: They explode! However,
besides this condition on the proton column, the instability imposes
a kinematic constraint on the synchrotron photon energy because the
synchrotron photons emitted by the $e^+e^-$--pairs must be able to
pair-produce in collisions with the protons. This effectively
requires the RBW Lorentz factor to be larger than a critical value,
$\Gamma_c$, given by the condition {$\Gamma_c^5 \, b \simeq 1$}, a
demand imposed by the kinematic threshold of the above reaction ($b
= B/B_{\rm cr}$ is the postshock value of the GRB magnetic field
with $B_{\rm cr} = 4.4 \times 10^{13}$ G the value of the critical
magnetic field). Incidentally, the energy of peak emission of the
prompt phase of this model, {\em on the Earth frame}, is also (in
units of $m_e c^2$) \Ep$\simeq\Gamma^5 b \simeq 1$, so in this
model, the observed characteristic value of \Eps reflects simply the
threshold energy of pair production on the GRB frame.

Another crucial element of this model is the upstream scattering of
the RBW synchrotron radiation and its re-interception by it. As
noted in \cite{mas08,mas09} and more specifically in \cite{SKM13},
this process induces a radiation reaction on the RBW and causes a
(relatively) small ($\sim 30\%-50\%)$ reduction of its Lorentz
factor over a radius $\Delta R \ll R$. Even though small, this
reduction is important because it pushes $\Gamma$ below the
threshold value $\Gamma_c$, thus arresting the transfer of energy
from the RBW relativistic protons into $e^+e^-$--pairs. As noted in
\cite{SKM13} this event should result in an abrupt reduction of the
RBW radiative flux (producing the observed steep decline), thereby
defining the end of the prompt GRB phase and the beginning of the
afterglow. The authors of \cite{SKM13} estimated the reduction in
flux to be by a factor roughly equal to the proton-to-electron mass
ratio, i.e. by $\sim m_p/m_e \simeq 2000$, since the emitted
radiation now comes from the cooling of {\em only
the electrons} being swept-up by the RBW. 
Furthermore, the radius at which this small, sharp reduction in
$\Gamma$ takes place is generally smaller than $R_D$. Because an RBW
expands with constant $\Gamma$ until it amasses enough inertia to
begin its decline at  $R > R_D$, it will continue to do so at its
new, smaller $\Gamma$, and with the GRB now being in its afterglow
stage. Then, depending on the rate of decrease of the RBW magnetic
field with radius or the density profile of the ambient medium
\cite{Matzner12}, the ensuing synchrotron (or Compton) emission
could be constant, decreasing slightly or even increasing with time,
producing thus the afterglow plateau emission. Finally, beyond the
deceleration radius, which is reached at time $T_{\rm brk}$ (a time
associated within the model with $R_D$) a more conventional decrease
in afterglow flux ensues, consistent with the standard
analyses\cite{sar98}.

Motivated by the specific value of the flux ratio, ${\mathrm R}$,
between the prompt GRB and the afterglow plateau stages implied by
the model put forward in \cite{SKM13}, we have compiled and present
in this note the distribution of ${\mathrm R}$ obtained from a large
number of GRB afterglows of the \Sw-{\em XRT} repository
\cite{evans09,evans10}. This is presented in the next section. Along
with this we also present a correlation between $L_{\rm iso}$, {the
peak isotropic luminosity} of the prompt emission and the afterglow
X-ray luminosity $L_X$ at time $t = T_{\rm brk}$, the time the X-ray
afterglow resumes its more conventional {decay}. We finish in \S 4
with our conclusions and some discussion.

\begin{figure}[t]
\begin{center}$
\begin{array}{cc}
\includegraphics[trim=0in 0in 0in 0in,keepaspectratio=false,width=3.0in,
angle=-0,clip=false]{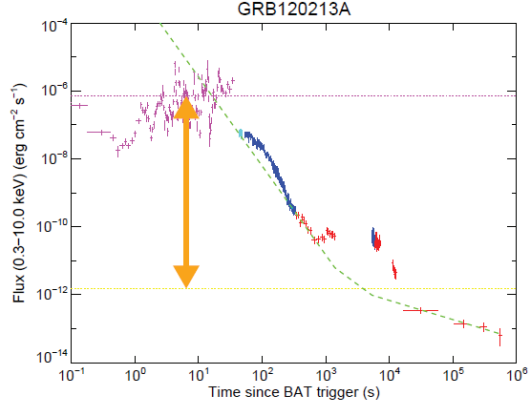}
\end{array}$
\end{center}
\vskip -20pt
\caption{\footnotesize  Same as in figure 1 but for GRB 120213A.
The two horizontal lines indicate the flux levels that our algorithm has chosen in
computing the ratio ${\mathrm R}$. Eye inspection indicates that a more reasonable value is
$\sim 10^4$, considering that there appears to be a plateau beginning at $t \simeq
10^3$ sec.}
\label{fig:f2}
\end{figure}

\section{The Prompt-to-Afterglow Flux Ratios}

Following the arguments presented above, we have searched the \Ss
data base and compiled the ratios, ${\mathrm R}$, between the prompt
and afterglow GRB fluxes.
We have used fits to the average BAT light curves obtained from the
\Sw--{XRT} repository Burst Analyzer \cite{evans10} extrapolated
down to the 0.3--10 keV band from fits to the { XRT}, as described
in \cite{racusin09,racusin11}, to extract the flux at the transition
between steep decline and plateau in the afterglows demonstrating
that form.
In figures 1a,b and 2 we present three specific cases of GRB
transitions from the prompt to the afterglow stages with a variety
of post transition behaviors. The fast decline exponents range
between $\alpha \simeq -3$ and $\alpha \simeq -6$, while their
transition to $\alpha \simeq -1$ happens in all cases around $T_{\rm
brk} \simeq 10^4$ sec. The yellow arrows show the decrease in flux
between the {\em geometric mean} of the highly variable prompt emission
and the end of the steep decline phase.

These figures indicate that these transitions have peculiarities of
their own. For example the decrease of GRB 050713A is by a factor
${\mathrm R} \simeq 10^3$, smaller than the $m_p/m_e$ ratio;
however, in the other two GRB the decrease, as shown by the yellow
arrows, is by factor $10^4$ and $10^6$, one of them significantly
larger than the $m_p/m_e$ ratio. On the other hand this last case,
namely of GRB 120213A, exhibits a rather peculiar two step
transition; in the first step, the flux decreases by $\sim 10^4$ and
it is followed by another one by an additional factor of $10^2$. The
horizontal lines and the yellow arrow in Fig. 2 indicate the fluxes
considered by the algorithm employed for extracting the afterglow
parameters. This last case indicates that, besides applying a given
algorithm, one may have to scrutinize each such transition
individually.

From the point of view of the data available in search of
correlations among GRB attributes, the one proposed in \cite{SKM13}
and tested herein has the clear advantage that it involves only flux
ratios rather than absolute values (whether luminosities, time lags
or values of \Ep) as is the case with many of the GRB produced
correlations (e.g. the Lag-Luminosity relation, the \Ep--E$_{\rm
iso}$,  the $L_X - T_{\rm brk}$. etc. correlations). As such,
knowledge of the GRB redshift is not necessary, a fact that allows
the compilation of a large number of bursts. The ratios of the
prompt to afterglow fluxes were compiled from the {\em Swift}--XRT
repository and spans the period between December 2004 and March
2014. 

The main result of our analysis is given in Fig. 3a where we present
a histogram of the logarithm of the BAT-to-XRT flux ratio, ${\mathrm
R}$, computed as described above along with a dashed vertical line
that indicates the value of the $m_p/m_e$ ratio. The distribution
exhibits a broad maximum at almost precisely this value, indicating
the presence of a characteristic ratio between the prompt and
afterglow fluxes, as proposed in \cite{SKM13}. The ${\mathrm
R}$-distribution appears to be log-normal, though its precise shape
is not easy to determine accurately. It spans 5-6 decades in
${\mathrm R}$, with a FWHM of about 2 decades and a median value of
$\log {\mathrm R}$ essentially equal to that of $\log(m_p/m_e)
\simeq 3.25$ and a slightly larger medium value ($\simeq 10^4$) in
sufficient agreement with the suggestion of \cite{SKM13} to merit
further consideration.

\begin{figure}[t]
\begin{center}$
\begin{array}{cc}
\includegraphics[trim=0in 0in 0in
0in,keepaspectratio=false,width=3.3in,angle=-0,clip=false]
{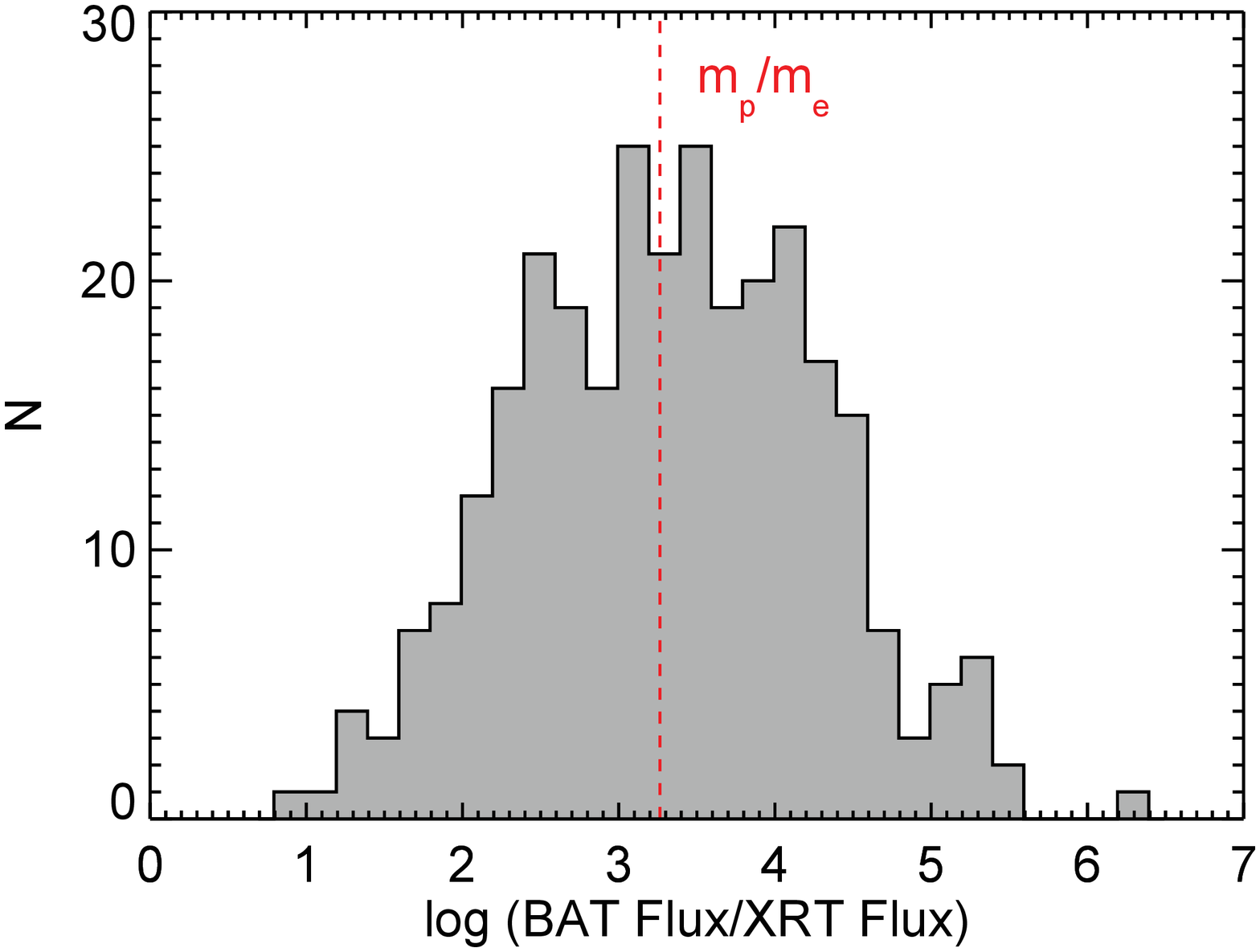} &
\includegraphics[trim=0in 0in 0in
0in,keepaspectratio=false,width=3.1in,angle=-0,clip=false]
{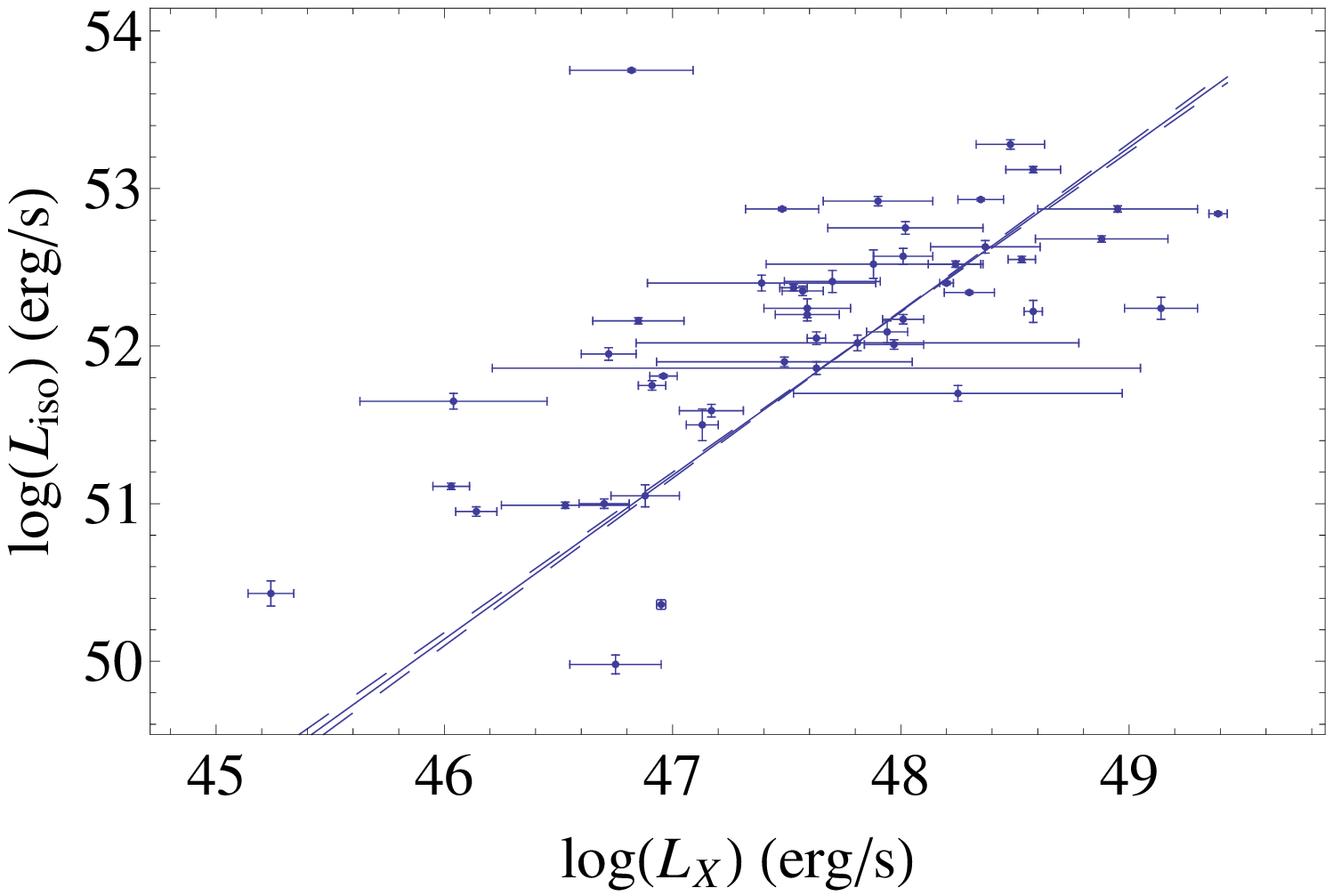}
\end{array}$
\end{center}
\vskip -20pt
\caption{\footnotesize  The histogram of the BAT to XRT flux ratio for a number of
{\em Swift} GRB. The distribution shows clearly a preferred value for this ratio of order
$\sim 10^3 - 10^4$. The vertical line shows also the proton to electron mass ratio $m_p/m_e$. }
\label{fig:f3}
\end{figure}

As mentioned above, a number of correlations has already been
established between GRB prompt or afterglow emission
characteristics, such as the Lag--Luminosity relation
\cite{nor00,nor02,S07}, the maximum prompt emission (isotropic)
luminosity $L_{\rm iso}$ and the peak energy of the Band function
$E_{\rm p}$ \cite{S07,WQD11} and the afterglow plateau X-ray
luminosity $L_{X}$ and the rest-frame plateau end-time $T_{\rm
brk}$, beyond which the afterglow resumes the standard decline
\cite{dai08,dai10,dai13}. In \cite{SK12} we showed that the
Lag--Luminosity of the prompt emission extrapolates into the
$L_{X}-T_{\rm brk}$ afterglow correlation, suggesting the intimate
connection of these phases, despite the apparent absence of
continuity between them (however, there is a correlation between the
two within the SPM). Motivated by this relation and the histogram of
Fig. 3a, we bypass the time coordinate of the relation given in
\cite{SK12} and plot in Fig. 3b the maximum prompt isotropic
luminosity $L_{\rm iso}$ vs. the X-ray luminosity of the afterglow
plateau segment at the time $T_{\rm brk}$. There appears to be a
correlation between these quantities. A least squares fit gives the
following relation between $L_{\rm iso}$ and $L_X$
\begin{equation}
\log L_{\rm iso} = (4.04 \pm 0.10) + (1.04 \pm 0.02) \log L_X
\end{equation}
with correlation coefficient $\rho = 0.69$. The ratio of these two
quantities appears consistent with that shown in Fig. 3a. Given the
difference in the choice of these samples and the slightly different
properties they depict, they appear to be consistent with each other
and the general premise of the prompt to afterglow luminosity
ratios.

\section{Discussion, Conclusions}

Motivated by the considerations put forward in \cite{SKM13}, based
on the SPM of GRB dissipation, we have compiled the flux ratios
between the prompt and the afterglow plateau stages of GRB to
indicate that there is indeed a characteristic value for the ratio
${\mathrm R}$ of these two quantities. The important point to bear
in mind is that this characteristic value for ${\mathrm R}$, namely
$m_p/m_e$, was proposed in \cite{SKM13} {\em before} the compilation
of the histogram of Fig. 3a; as such, it constitutes a {\em
prediction} of the model, one of the very few in the GRB field of
study. A similar relation has also been found between slightly
different quantities of these two GRB phases, shown in Fig. 3b, one
that requires, however, knowledge of their redshifts.

One must note at this point that though there is a maximum in the
distribution of fluxes near the value $m_p/m_e$, the histogram of
Fig.3 has a finite width. Thus there are bursts with ${\mathrm R}$
values as large as $10^6$ and as low as $10^2$. Figure 2 shows a
burst with a particularly large value of this ratio. As argued
earlier, one could assign to this burst a value smaller than that
given by our algorithm, given the peculiar form of its afterglow. On
the other hand, if one takes into account that the observed
luminosity has a dependence on the Lorentz factor of the flow
$\Gamma$ as strong as $\Gamma^4$, even a small reduction in $\Gamma$
could increase the pre-to-post prompt emission fluxes to values
larger than $m_p/m_e$. Values of ${\mathrm R} < m_p/m_e$ would
appear to be more problematic. An account of these values, put
forward in \cite{SKM13}, is that not all protons "are burnt" in
prompt phase, thus reducing the flux of this stage. A different
possibility is that in these cases, the original angle of the jet to
the observer's line of sight, $\theta$, is slightly larger than
$1/\Gamma$, yielding a reduced relativistic boosting for the prompt
emission; after the RBW radiation-reaction slowdown, the
smaller value of $\Gamma$ allows the observer's line of sight to
"peer" directly into the (wider now)
relativistic outflow, thereby reducing the ratio of the pre-to-post
prompt emission fluxes. If this is the case, then the prompt
emission of bursts with $R < m_p/m_e$ should be indicative of this
situation, e.g. they should exhibit longer Lags, smaller \Ep,
smaller L$_{\rm iso}$, issues that could in principle be tested by
an appropriate choice of a GRB sample. However, such considerations
are beyond the scope of the present note.

\end{document}